\documentclass[aps,prl,reprint,showpacs,showkeys,superscriptaddress,preprintnumbers,nofootinbib]{revtex4-1}
\pdfoutput=1
\usepackage{amsmath,amssymb,bm,mathrsfs}
\usepackage{hyperref}
\usepackage{url}
\usepackage{breakurl}
\usepackage{graphicx}

\usepackage[caption=false]{subfig}

\usepackage{comment}

\begin{document}


\preprint{DESY 15-054}
\preprint{IPMU-15-0053}

\title{Wino dark matter in light of the AMS-02 2015 data}

\author{Masahiro Ibe}
\affiliation{Institute for Cosmic Ray Research (ICRR), Theory Group, University of Tokyo, Kashiwa, Chiba 277-8568, Japan}
\affiliation{Kavli Institute for the Physics and Mathematics of the Universe (IPMU),University of Tokyo, Kashiwa, Chiba 277-8568, Japan} 
\author{Shigeki Matsumoto}
\affiliation{Kavli Institute for the Physics and Mathematics of the Universe (IPMU),University of Tokyo, Kashiwa, Chiba 277-8568, Japan} 
\author{Satoshi Shirai}
\affiliation{Deutsches Elektronen-Synchrotron (DESY), 22607 Hamburg, Germany}
\author{Tsutomu T. Yanagida}
\affiliation{Kavli Institute for the Physics and Mathematics of the Universe (IPMU),University of Tokyo, Kashiwa, Chiba 277-8568, Japan} 
\begin{abstract}
The AMS-02 collaboration has recently reported the antiproton to proton ratio with improved accuracy.
In view of uncertainties of the production and the propagation of the cosmic rays,
the observed ratio is still consistent with the secondary astrophysical antiproton to proton ratio. 
However, it is nonetheless enticing to examine whether the observed spectrum can be explained by
a strongly motivated dark matter, the wino dark matter.
As we will show, the antiproton flux from the wino annihilation can explain the observed 
spectrum well for its mass range 2.5--3\,TeV.
The fit to data becomes particularly well compared to the case without the annihilation
for the thermal wino dark matter case with a mass about 3\,TeV. 
The ratio is predicted to decrease quickly at the energy several hundreds of GeV,
which will be confirmed or ruled out in near future 
 when the AMS-02 experiment accumulates enough data at this higher energy region.
\end{abstract}

\maketitle

\setcounter{footnote}{0}

\section{Introduction}
The AMS-02 collaboration has recently reported the antiproton to proton ratio with improved 
accuracy\,\cite{AMS02:pbar_p}. 
The observed spectrum of the antiproton fraction looks flatter 
than the one expected for the secondary astrophysical antiproton.
At this point, however, it is premature to say that the observed fraction requires new sources of antiproton
such as dark matter.
In fact, the detailed analyses in \cite{Giesen:2015ufa,Jin:2015sqa} have shown 
that the observed spectrum is still consistent with the 
one of the secondary antiproton within the uncertainties of the production and the propagation 
of the cosmic rays.

Having said so, it is nonetheless  enticing to examine whether a theoretically motivated dark matter candidate,
the wino dark matter, can fit the spectrum when the secondary astrophysical antiproton cannot fully explain 
the spectrum of the fraction.
The wino dark matter is, in fact, anticipated in a wide class of supersymmetric standard models
where the gaugino masses are generated by the anomaly mediated supersymmetry breaking contributions\,\cite{AMSB}. 
In particular, in conjunction with the high scale supersymmetry breaking\,\cite{Wells:2004di,Ibe:2006de,PGMs, PGM2, Evans:2013lpa,Strong Moduli,spread} (see also \cite{Split SUSY}), the models with anomaly mediated gaugino mass are considered to be one of the most attractive possibilities.
In addition to a good dark matter candidate (i.e. the wino), this class of models  
explains the observed Higgs boson mass about 125\,GeV\,\cite{OYY}
simultaneously.%
\footnote{Apart from the supersymmetric theories, the wino-like dark matter is also discussed extensively
as ``minimal dark matter scenario"~\cite{Cirelli:2005uq} (see also \cite{Ibe:2009gt}). 
}

As a phenomenologically notable feature of the wino dark matter,
it is not only a good candidate for weakly interacting massive particle (WIMP) but also predicts 
a rather large annihilation cross section (mainly into a $W^\pm$ pair)
due to the so-called Sommerfeld enhancement\,\cite{Hisano:2003ec, Hisano:2004ds,Hisano:2005ec}.
Thus, the wino dark matter predicts strong signals in indirect detection searches.
In particular, the signals in the antiproton flux is one of the promising discovery channels 
of the wino dark matter, where the antiprotons are produced from the $W^\pm$ in the main 
annihilation mode.

In this paper, we demonstrate how well the observed spectrum of the antiproton fraction can be fitted 
by the annihilation of the wino dark matter.
As we will show,  the antiproton flux from the wino annihilation can explain the observed 
spectrum very well for its mass about $2.5$--$3$\,TeV.
In particular, the fit to the data  becomes very well compared to the case without the annihilation
for the thermal wino dark matter case with a mass about $3$\,TeV. 
It should be emphasized  that the wino dark matter has only one free parameter, the mass of the wino $M_{\tilde w}$,
and hence, it is quite non-trivial that the spectrum can be fitted very well by the wino annihilation contribution.

\section{The Wino  Dark Matter}
Let us briefly review the wino dark matter in the high scale supersymmetry breaking models.
Here, we take the pure gravity mediation model\,\cite{PGMs, PGM2} as an example, 
although the following properties are not changed significantly in other models
as long as the Higgsinos are heavy enough.
In the model, the gaugino masses are dominantly generated by the anomaly mediation\,\cite{AMSB},%
\footnote{See \,\cite{Bagger:1999rd, D'Eramo:2013mya, Harigaya:2014sfa}
for discussion of the anomaly mediation mechanism in superspace formalism of supergravity.}
which are one-loop suppressed compared to the gravitino mass.
The Higgsino mass is, on the other hand, generated via tree-level interactions to the $R$-symmetry breaking sector\,\cite{Casas:1992mk} (see \cite{Giudice:1988yz} for a related mechanism), 
which leads to a much heavier Higgsino than the gauginos.
As a result, the pure gravity mediation model predicts the almost pure neutral wino as the lightest supersymmetry particle (LSP)
which is a good candidate for WIMP dark matter.

As mentioned earlier, the wino dark matter possesses a phenomenologically notable feature, a large 
annihilation cross section enhanced by the so-called Sommerfeld effects\,\cite{Hisano:2003ec, Hisano:2004ds,Hisano:2005ec}.
Due to the enhancement, the annihilation cross section into a pair of the $W$-bosons at present universe 
is automatically boosted to be $10^{-24}$--$10^{-25}$\,cm$^3$/s. 
In Fig.~\ref{subfig:cons}, we show the annihilation cross section of the wino dark matter into a pair of $W$ bosons as a solid line.
With this large cross section, the antiproton flux from the wino annihilation can be 
comparable to the secondary astrophysical antiproton flux at $T_p \gtrsim 100$\,GeV, with $T_p$ being the kinetic energy of a proton and an antiproton.

There are two favored mass regions for the wino dark matter.
One is the mass region around 3\,TeV where the observed dark matter density is explained solely
by its thermal relic density\,\cite{Hisano:2006nn}.
The other region is below  $1$--$1.5$\,TeV where the relic density is provided non-thermally 
by the decay of the gravitino\,\cite{Gherghetta:1999sw,Moroi:1999zb}.
There, the appropriate gravitino abundance for the non-thermal wino production 
is achieved  when the reheating temperature of the universe is consistent 
with the traditional thermal leptogenesis scenario\,\cite{leptogenesis}.
As we will see shortly, the wino mass in the both mass regions can sizably contribute to
the antiproton spectrum, although the thermal wino case fits the observed spectrum of the antiproton fraction particularly well.

So far, the mass of the wino dark matter has been constrained by collider experiments.
Among them, the searches for disappearing tracks made by a short lived charged wino 
inside the detectors put a lower limit on the mass of the wino dark matter,
\begin{eqnarray}
M_{\tilde w} \gtrsim 270\,{\rm GeV}\ ,
\end{eqnarray}
with $20$\,fb$^{-1}$ data at LHC 8\,TeV running~\cite{Aad:2013yna}.%
\footnote{See \cite{Ibe:2012sx} for a two-loop calculation of the wino mass splitting.}
At the 14\,TeV running, the limit can be pushed up to $500$\,GeV with $100$\,fb$^{-1}$ data~\cite{Yamanaka}.
See also Refs.~\cite{Cirelli:2014dsa,Low:2014cba,Harigaya:2015yaa} for more details on the future prospects of the wino dark matter
searches at the collider experiments.

The wino dark matter is also constrained by the indirect detections of dark matter in cosmic-rays.
To date, the most robust limit comes from the gamma-ray searches from the dwarf spheroidal galaxies (dSphs) 
at the Fermi-LAT experiment. 
By taking uncertainties of the dark matter profile of the dSphs,
it has excluded $M_{\tilde w} \lesssim 320$\,GeV and
$2.25$\,TeV$\lesssim M_{\tilde w} \lesssim 2.43$\,TeV 
at the 95\% confidence level (C.L.) using four-year data~\cite{Ackermann:2013yva}.%
\footnote{For uncertainties and future prospects of 
the searches for the wino dark matter via the gamma-rays from the dSphs, see 
e.g.~\cite{Bhattacherjee:2014dya,Geringer-Sameth:2014qqa}. }
It should be noted that the constraints on the wino dark matter via monochromatic gamma-ray searches
from the galactic center~\cite{Abramowski:2013ax} and from the dSphs~\cite{Abramowski:2014tra}
by the H.E.S.S experiments are less stringent due to
large uncertainties of the dark matter profile at the galaxy center (see e.g. Ref.\,\cite{Nesti:2013uwa}) 
and the small cross section into the monochromatic gamma-rays.

\section{Antiproton Flux from the Wino Annihilation }
\begin{figure*}[t]
\subfloat[Constraints]{
 \label{subfig:cons}
 \includegraphics[width=0.49\textwidth]{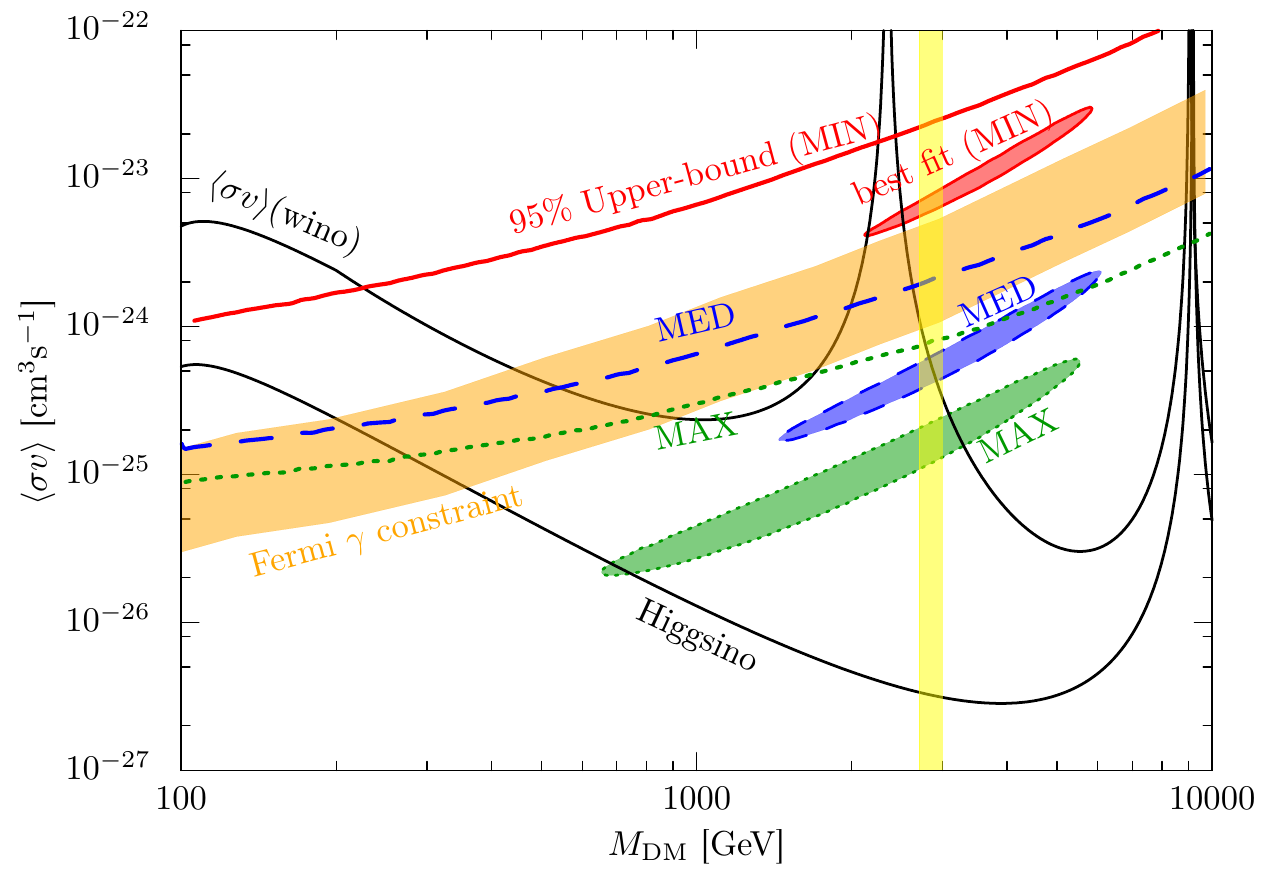}
}
\subfloat[Antiproton to proton ratio]{
 \label{subfig:ratio}
 \includegraphics[width=0.49\textwidth]{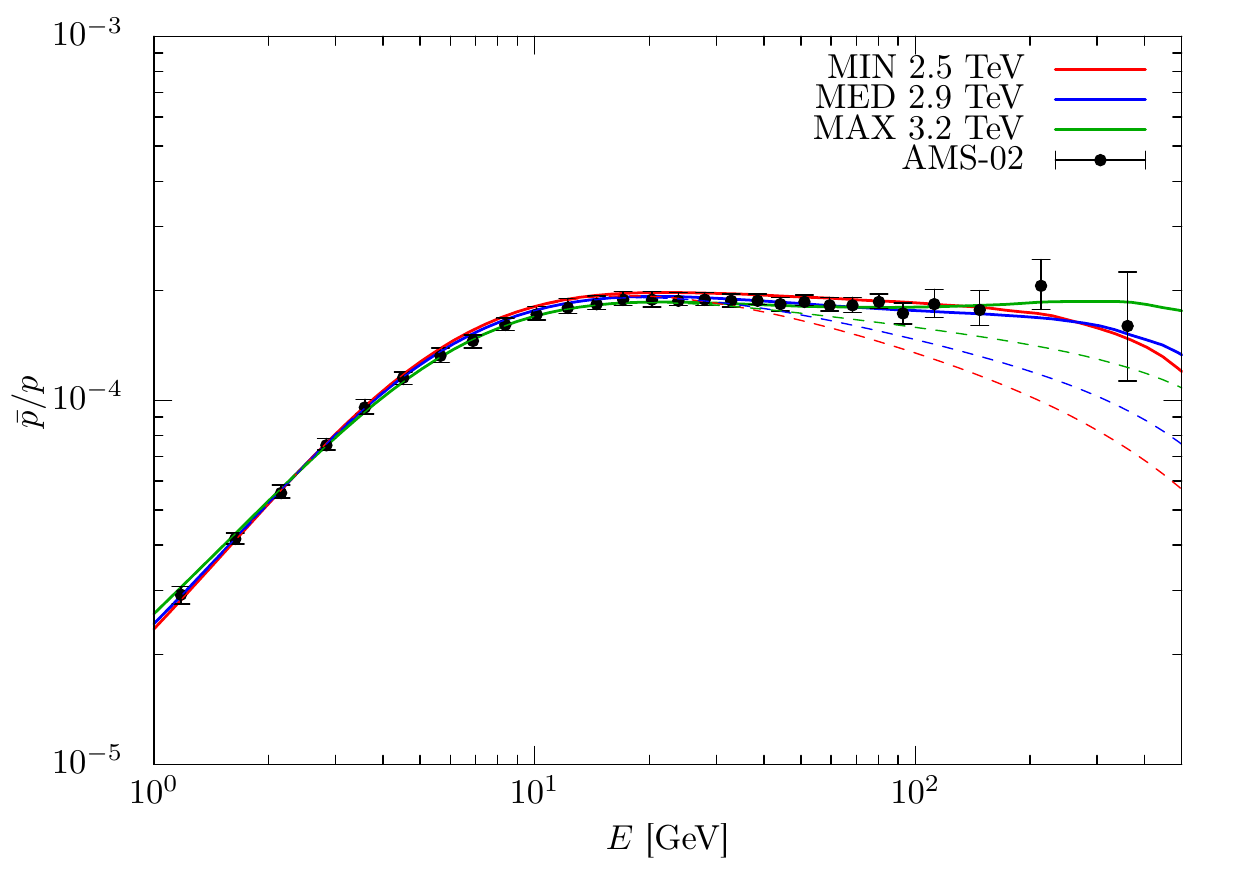}
}
\caption{(a): Constraints on the ($M_\mathrm{DM}$-$\langle \sigma v\rangle$) plane.
The black solid lines show the predicted annihilation cross sections for the wino and Higgsino.
Red solid, blue dashed and green dotted lines show the  upper-bounds on the annihilation cross section at 95\% C.L. for MIN, MED and MAX propagation models, respectively.
The shaded regions with same color show the best-fitted regions.  
The constraint from the Fermi is shown with the orange bands.
The yellow vertical shaded region indicates the wino mass range where the wino thermal relic abundance is the observed dark matter density. 
(b): Predicted antiproton to proton ratio with experimental data.
The solid (dashed) lines show the case with (without) the dark matter contributions.
}
\label{fig:fit}
\end{figure*}

Now, let us discuss the antiproton flux from the annihilation of the wino dark matter.
The wino annihilation in the dark matter halo produces the weak bosons, whose subsequent decay and hadronization make the antiprotons.
In this work, we assume the dark matter mass density is the NFW profile\,\cite{Navarro:1996gj} with profile parameters $\rho_\odot = 0.4~{\rm GeV/cm^3}$ (the local halo density), $r_c = 20~{\rm kpc}$ (the core radius), and $r_\odot = 8.5~{\rm kpc}$.%
\footnote{In the case of the NFW profile, the gamma-ray constraints from the galactic center are severe.
However, these constraints strongly owe to the assumption that the dark matter density profile at the galactic center exactly obeys the NFW profile.
A small modification of the central structure can drastically relax the gamma-ray constraints, while the effect to the antiproton flux is small.
}
The antiproton energy spectrum is estimated with the program {\tt PYTHIA6}~\cite{Sjostrand:2006za}.
We have used the programs {\tt DRAGON}~\cite{Evoli:2008dv}, to calculate the antiproton 
propagation in the galaxy.

In Fig.~\ref{subfig:cons}, we show the constraints on the parameters $M_\mathrm{DM}$ and $\langle \sigma v\rangle$.
The red solid, blue dashed, and green dotted lines show the upper-bound on the annihilation cross section at 95\% C.L. for MIN, MED and MAX propagation models~\cite{Donato:2003xg}, respectively.
To get the conservative upper-bounds, we assume the background antiproton spectrum is arbitrary.
The figure shows, for example, that $m_{\tilde w} \lesssim 500$\,GeV is excluded for MED propagation model,
although the constraint is much weaker for MIN propagation model.

In the figure, we also show the preferred parameter space as the shaded regions 
for each propagation model to explain the AMS-02 ``excess" (1$\sigma$ level).
In this analysis, we take the best-fitted background of Ref.~\cite{Giesen:2015ufa} assuming 
the background only hypothesis,  and add the dark matter contributions.
For the fitting, we use the AMS-02 $\bar{p}/p$ data with $T_p>50$ GeV.

The orange band shows the upper-bound on the annihilation cross section from dSphs with six years Fermi-LAT data~\cite{Ackermann:2015zua}.
Here, the width of the band represents an uncertainty of the constraint from the ultra-faint dSphs
which comes from the uncertainties of the dark matter density profile of the ultra-faint dSphs.
According to Ref.~\cite{Ackermann:2015zua}, we adopt a factor $5$ as an uncertainty of the Fermi-LAT constraints. 
We show the cross section of the wino and  the Higgsino annihilation to the weak bosons (upper and lower black solid line, respectively). 
The yellow vertical band shows the mass range in which the thermal relic abundance of the wino is the observed dark matter density~\cite{Hisano:2006nn}.

In Fig.~\ref{subfig:ratio}, we show the antiproton to proton ratio with the wino dark matter contributions.
Here we take the wino mass 2.5 TeV for MIN (red), 2.9 TeV for MED (blue) and 3.2 TeV for MAX  (green), 
which give the best fits.
The dashed lines show the best fit result without the dark matter contributions~\cite{Giesen:2015ufa}.

Note that the estimation of the antiproton flux have various uncertainties~\cite{Evoli:2011id}.
The most important uncertainty is the propagation model as seen in Fig.~\ref{subfig:cons}.
Another significant effect comes from the dark matter halo model.
For instance, if we adopt the  Burkert halo profile~\cite{Burkert:1995yz}, a few times larger cross section is needed for the 
best fit, depending on the propagation model.
The uncertainty of the local dark matter density also affects the predicted cross section, which is scaled as $(\rho_\odot/ 0.4~{\rm GeV \cdot cm^{-3}})^{-2}$.
The higher order corrections  to the annihilation process \cite{Ciafaloni:2010ti,Hryczuk:2011vi} and uncertainty of the hadronization affect the prediction of the antiproton flux by $O(10)$\%.

In this analysis, we have fixed astrophysical backgrounds to the best-fitted ones  of Ref.~\cite{Giesen:2015ufa}.
Let us here comment on the case that we fully fit the spectrum with both the background and the dark matter contributions.
In the low-energy region $T_p<O(10)$ GeV, the contributions from the dark matter get tiny and the antiproton to proton ratio in this region determined almost solely by the background contributions.
For the higher energy region, on the other hand, both the background and dark matter contributions
are comparable for the background we took in our analysis.
Thus, for a smaller background antiproton flux, the larger dark matter contributions are required to compensate the spectrum.
Therefore, we expect that full fitting (including background flux) leads to a larger best fit region towards a larger cross section and a smaller mass region, so that the dark matter contribution can be enhanced.

With these uncertainties, it is hard to conclude that the AMS-02 result points only the 3 TeV wino region.
Depending on these uncertainties, the lighter wino ($M_{\tilde w}\lesssim 1.5$ TeV), can also fit the  antiproton to proton ratio, if the non-thermal wino production realizes the observed dark matter density. 
For instance, the 1.5 TeV wino with the MED propagation model and the lower local halo density e.g., $0.3~\rm{GeV\cdot cm^{-3}}$ also provides good fitting, as seen in Fig.~\ref{subfig:cons}.
However we expect the 3 TeV region is always preferred, even if we include these uncertainties.

\section{Summary and discussions}
We have examined how the annihilation of the wino dark matter affects the antiproton to proton ratio in the light of new data reported by the AMS-02 collaboration\,\cite{AMS02:pbar_p}.
As a result, we found that the annihilation of the wino dark matter can explain the observed spectrum of the antiproton
fraction  for $M_{\tilde w}  \simeq 2.5-3$\,TeV  when the spectrum cannot be 
fully explained by the secondary astrophysical antiprotons.
The fit to the data becomes particularly well compared to the case without the annihilation
for the thermal wino dark matter, i.e. $M_{\tilde w} \simeq 3$\,TeV
as can be clearly seen in Fig.~\ref{fig:fit}.
It is worth notifying that the lighter wino ($M_{\tilde w}\lesssim 1.5$ TeV) can also
account for the AMS-02 excess because of several uncertainties
on the propagation of antiprotons, the DM profile, etc.,
as mentioned before.

It is of course premature to conclude that the wino dark matter is needed to explain the ratio reported by the AMS-02 collaboration. The observed data is still consistent with the traditional secondary astrophysical antiproton to proton ratio within systematic uncertainties associated with cosmic-ray propagation\,\cite{Giesen:2015ufa, Jin:2015sqa}. However, this interesting possibility of the wino contribution can be tested in near future when the AMS-02 experiment accumulates more data on the ratio at $T_p \gg O(100)$\,GeV, for the constitution is predicted to be decreased quickly at this $T_p$ region.

In addition to the antiproton flux, the AMS-02 can also precisely measure 
other secondary-to-primary ratios such as boron-to-carbon (B/C), which will lead to very strong constraints on the cosmic-ray propagation model~\cite{Pato:2010ih}.
This high-precision measurement may reveal the wino dark matter really account for the AMS-02 ``anomaly."

Several comments are in order.
Besides the antiproton to proton ratio, the AMS-02 collaboration has also reported the electron and the positron fluxes as well as the positron fraction with high accuracy\,\cite{Aguilar:2013qda, Accardo:2014lma, Aguilar:2014mma}. As is well known, these data seem to require some new contributions to those fluxes in addition to the standard ones. 
The annihilation of the wino dark matter (without any astrophysical boost factors unfortunately) 
gives too small contributions to the fluxes when its mass is $O(1)$\,TeV\,\cite{Kopp:2013eka}.
Thus, the anomalies should be explained by some other sources, 
such as nearby pulsars\,\cite{Grasso:2009ma, Hooper:2008kg, Delahaye:2010ji, Linden:2013mqa}
in the case that the antiproton flux is explained by the wino annihilation.
As an alternative possibility, it is also possible to explain the excesses in the positron flux/fraction
by the decay of the wino dark matter with a lifetime of $10^{26}$--$10^{28}$ sec
caused by a slight $R$-parity violation by $LLE^c$-type interactions\,\cite{Shirai:2009fq,Ibe:2013jya, Ibe:2014qya}.
With the decay, the wino with a mass of around $3$~TeV can explain the observed positron/electron spectrum.
In this model, the decay does not contribute to the antiproton flux significantly.
Therefore the decaying wino dark matter can explain the observed antiproton and positron fluxes simultaneously.

\vspace{0.5cm}
\noindent
{\bf Acknowledgments}
\vspace{0.1cm}\\
\noindent
This work is supported by the Grant-in-Aid for Scientific research from the Ministry of Education, Science, Sports, and Culture (MEXT), Japan and from the Japan Society for the Promotion of Science (JSPS), Nos. 24740151 and 25105011 (M.I.), No. 26287039 (M.I., S.M. and T.T.Y.) and No. 26287039 (S.M. and T.T.Y.), as well as by the World Premier International Research Center Initiative (WPI), MEXT, Japan (M.I., S.M. and T.T.Y.). S.M. also thanks Koji Ichikawa for fruitful discussions about present constraints on the wino dark matter.


\bibliographystyle{aps}

\end{document}